\documentclass[12pt, preprint]{aastex}

\usepackage{epsfig}
\usepackage{url}
\usepackage{multirow}

\newcommand{\rmsub}[1]{\ensuremath{_{\mathrm{#1}}}}
\newcommand{\tento}[1]{\ensuremath{\times 10^{#1}}}
\newcommand{\us}{\ensuremath\mathrm{\mu s}} 
\newcommand{\ms}{\textrm{ms}} 
\newcommand{\s}{\textrm{s}} 
\newcommand{\yr}{\textrm{yr}} 
\newcommand{\ps}{\ensuremath{\mathrm{s^{-1}}}} 
\newcommand{\Myr}{\textrm{Myr}} 
\newcommand{\Gyr}{\textrm{Gyr}} 
\newcommand{\pssq}{\ensuremath{\mathrm{s^{-2}}}} 
\newcommand{\m}{\textrm{m}} 
\newcommand{\km}{\textrm{km}} 
\newcommand{\pc}{\textrm{pc}} 
\newcommand{\kpc}{\textrm{kpc}} 
\newcommand{\pcmsq}{\ensuremath{\mathrm{cm^{-2}}}} 
\newcommand{\pmsq}{\ensuremath{\mathrm{m^{-2}}}} 
\newcommand{\K}{\textrm{K}} 
\newcommand{\GHz}{\textrm{GHz}} 
\newcommand{\MHz}{\textrm{MHz}} 
\newcommand{\erg}{\textrm{erg}} 
\newcommand{\mJy}{\textrm{mJy}} 
\newcommand{\uJy}{\ensuremath{\mathrm{\mu Jy}}} 
\newcommand{\keV}{\textrm{keV}} 
\newcommand{\MeV}{\textrm{MeV}} 
\newcommand{\dmu}{\ensuremath{\mathrm{pc\, cm^{-3}}}}
\newcommand{\Msun}{\ensuremath{\mathrm{M_{\Sun}}}} 

\shorttitle{Six New Globular Cluster Pulsars} 
\shortauthors{Lynch et al.}

\begin{document}

\title{Six New Recycled Globular Cluster Pulsars Discovered with the
  Green Bank Telescope}

\author{Ryan S.\ Lynch\altaffilmark{1}, Scott M.\
  Ransom\altaffilmark{2}, Paulo C.\ C.\ Freire\altaffilmark{3}, and
  Ingrid H.\ Stairs\altaffilmark{4}} 
\altaffiltext{1}{Department of Astronomy, University of Virginia,
  P.O.\ Box 400325, Charlottesville, VA 22904-4325
  \texttt{rsl4v@virginia.edu}}
\altaffiltext{2}{National Radio Astronomy Observatory, 520 Edgemont
  Road, Charlottesville, VA 22903-4325, \texttt{sransom@nrao.edu}}
\altaffiltext{3}{Max-Planck-Institut f\"{u}r Radioastronomie, Auf dem
  H\"{u}gel 69, D-53121 Bonn, Germany,
  \texttt{pfreire@mpifr-bonn.mpg.de}}
\altaffiltext{4}{Department of Physics and Astronomy, University of
  British Columbia, 6224 Agricultural Road, Vancouver, BC V6T 1Z1,
  Canada, \texttt{stairs@astro.ubc.ca}}


\setcounter{footnote}{4}

\begin{abstract}

  We have completed sensitive searches for new pulsars in seven
  globular clusters using the Robert C.\ Byrd Green Bank Telescope,
  and have discovered six new recycled pulsars (four in NGC~6517 and
  two in M22), five of which are fully recycled millisecond pulsars
  with $P < 10\; \ms$.  We report full timing solutions for all six
  new pulsars and provide estimates of their flux densities and
  spectral indices.  None of the pulsars are detected in archival
  Chandra data down to $L\rmsub{X} \sim 10^{32}\; \erg\, \ps$ for
  NGC~6517 and $L\rmsub{X} \sim 10^{31}\; \erg\, \ps$ for M22.  One of
  the millisecond pulsars in M22 appears to have a very low mass
  companion, and is likely a new ``black widow''.  A second binary
  pulsar in NGC~6517 is in a long-period, mildly eccentric orbit.  We
  are able to set some lower limits on the age of the system, and find
  that it may be less than a few hundred million years old, which
  would indicate recent pulsar recycling in NGC~6517.  An isolated
  pulsar in NGC~6517 that lies about 20 core radii from the cluster
  center appears to have been ejected from the core by interacting
  with a massive binary.  By analyzing the luminosity function of the
  pulsars in NGC~6517, we predict the cluster to harbor roughly a
  dozen pulsars.  We use the observed period derivatives of three
  pulsars to set lower limits on the mass-to-light ratios in the cores
  of their host clusters, and find no evidence for large amounts of
  low-luminosity matter.  We also discuss reasons for non-detections
  in some of the clusters we searched.

\end{abstract}

\section{Introduction \label{sec:intro}}

The first globular cluster (GC) pulsar was discovered by
\citet{lbm+87}. Since then, 143 pulsars have been discovered in 27
GCs\footnote{For an up-to-date list see
  \url{http://www.naic.edu/~pfreire/GCpsr.html}}, the vast majority of
which are millisecond pulsars (MSPs).  In fact, over half of all known
MSPs are in clusters, which has been attributed to frequent dynamical
interactions that create mass-transferring binaries that are capable
of forming recycled pulsars \citep{cr05}, as well as to very deep,
targeted surveys.  These same dynamical encounters give rise to
systems that are seen only rarely, if at all, in the Galaxy, such as
the fastest spinning MSP \citep{hrs+06}, highly eccentric binaries
\citep{rhs+05,frg07}, massive neutron stars \citep{frb+08},
pulsar-main sequence binaries \citep{dpm+01}, and many ``black widow''
systems \citep{kbr+05}.  After a burst of activity in the early 1990s,
the pace of discovery of GC pulsars slowed until about 2000, after
which improvements in sensitivity and computing power led to an
explosion of new pulsars.  The Robert C.\ Byrd Green Bank Telescope
(GBT), completed in 2001, has been especially important, having
discovered 70 GC pulsars.  Despite this, most searches of GCs are
still sensitivity limited \citep{ran08}, meaning that we have only
begun to scratch the surface of this exciting population.

\citet{p+03} have shown that a good predictor of the number of low
mass X-ray binaries (LMXBs) in GCs is the two-body core interaction
rate, $\Gamma\rmsub{c} \propto \rho_0^{1.5} r\rmsub{c}^2$, where
$\rho_0$ is the central density and $r\rmsub{c}$ is the core radius.
As LMXBs are the progenitors to MSPs, one may expect that
$\Gamma\rmsub{c}$ would also be a good predictor of the number of
cluster pulsars, and this is indeed observed, especially when scaled
by the distance $D$ of the GC to account for flux losses (i.e.,
$\Gamma\rmsub{c}/D^2$).  This parameter was used to select twelve
promising clusters for GBT surveys in 2004--2005.  Results of these
surveys include the rich GCs Terzan 5 \citep{rhs+05}, M28 (B\'{e}gin
et al., in prep.), NGC~6440, and NGC6~441 \citep{frb+08}, all of which
have been shown to contain many pulsars.

Here we report on searches of an additional seven GCs that have
resulted in the discovery of six new pulsars.  In \S\ref{sec:searches}
we describe the sample of clusters that were targeted and the
parameters of our searches.  Follow-up timing observations of the
newly discovered pulsars are described in \S\ref{sec:timing} with
specific results presented in \S\ref{sec:results}.  We summarize our
results in \S\ref{sec:conc}.

\section{Search Parameters \label{sec:searches}}

Data were collected with the GBT at a central frequency of $2\; \GHz$
using $800\; \MHz$ of instantaneous bandwidth, although persistent
radio frequency interference (RFI) reduced the usable bandwidth to
$600\; \MHz$.  The Pulsar Spigot back-end \citep{k+05} was used in a
mode that offered 1024 frequency channels over $800\; \MHz$ of
bandwidth and a sampling time of $80.96\; \us$.  Total system
temperatures were typically $24$--$30\; \K$.  The contribution from
the Galactic background was estimated by scaling the values from
\citet{hss+82} to $2\; \GHz$ assuming a spectral index of $-2.6$, and
was usually $T\rmsub{Gal} \lesssim 2\; \K$ (though Liller 1 and Terzan
6, being closer to the Galactic plane, suffered from $T\rmsub{Gal}$ of
$7$ and $5\; \K$, respectively).  Integration times and approximate
limiting flux densities can be found in Table \ref{table:GCs}, along
with some other properties of each cluster.

Liller 1 was also searched in 2007, taking advantage of a new, 2048
frequency channel Spigot mode.  Liller 1 is an intriguing cluster
because despite a very high central density and the presence of
unresolved, steep spectrum radio emission in its core \citep{fg00}, no
pulsars have been detected.  The likely culprit is dispersive smearing
and scatter broadening of pulsar signals caused by free electrons
along the line of sight, thanks to its location very near the bulge of
the Galaxy (Liller 1 is predicted to have a dispersion measure,
$\mathrm{DM} \sim 740\; \dmu$).  The improved frequency resolution of
the new Spigot mode offered a factor of two improvement in dispersive
smearing compared to previous searches at $2\; \GHz$.  We also
observed Liller 1 at a frequency of $4.8\; \GHz$, hoping to overcome
scattering (which roughly scales as $f^{-4}$), while retaining
sensitivity to any bright pulsars (pulsars are steep spectrum sources
and thus dimmer at high frequency).

All searches were processed using the \texttt{PRESTO} software suite
\citep{rem02}.  After removing RFI, the data were transformed to the
Solar System barycenter, and de-dispersed time series were created at
a range of trial DMs that were based upon the NE2001 model of free
electron density \citep{cl02}.  Because of uncertainties in this
model, and drawing on past experience with similar pulsar searches, we
used a minimum DM of $10\; \dmu$ and maximum DM $1.5$ times greater
than that predicted by NE2001.  Acceleration searches for isolated and
binary pulsars were carried out in the Fourier domain for signals with
maximum accelerations of $z\rmsub{max} = \pm 400$--$800$ Fourier bins.
This corresponds to a physical acceleration of
\begin{eqnarray}
  a\rmsub{max} = \frac{z\rmsub{max} c P}{t\rmsub{obs}^2}
\end{eqnarray}
where $c$ is the speed of light, $P$ is the spin period of the pulsar,
and $t\rmsub{obs}$ is the length of the data that was searched.  We
searched the full length observations to maximize our sensitivity to
weak pulsars, and also broke the observations into 10 and 30 minute
chunks to increase sensitivity to bright pulsars in tight binaries.
Exact values of $a\rmsub{max}$ for full length searches for a $3\;
\ms$ pulsar can be found in Table \ref{table:GCs}.  We also carried
out single-pulse searches to look for transient emission.  Final
candidates were visually inspected and grouped into potential pulsars,
RFI, or random noise.  In many cases a GC was observed twice, in which
case we were quickly able to confirm or reject marginal candidates.

\section{Pulsar Timing Analysis \label{sec:timing}}

\begin{figure}
\centering 
\includegraphics[width=6in]{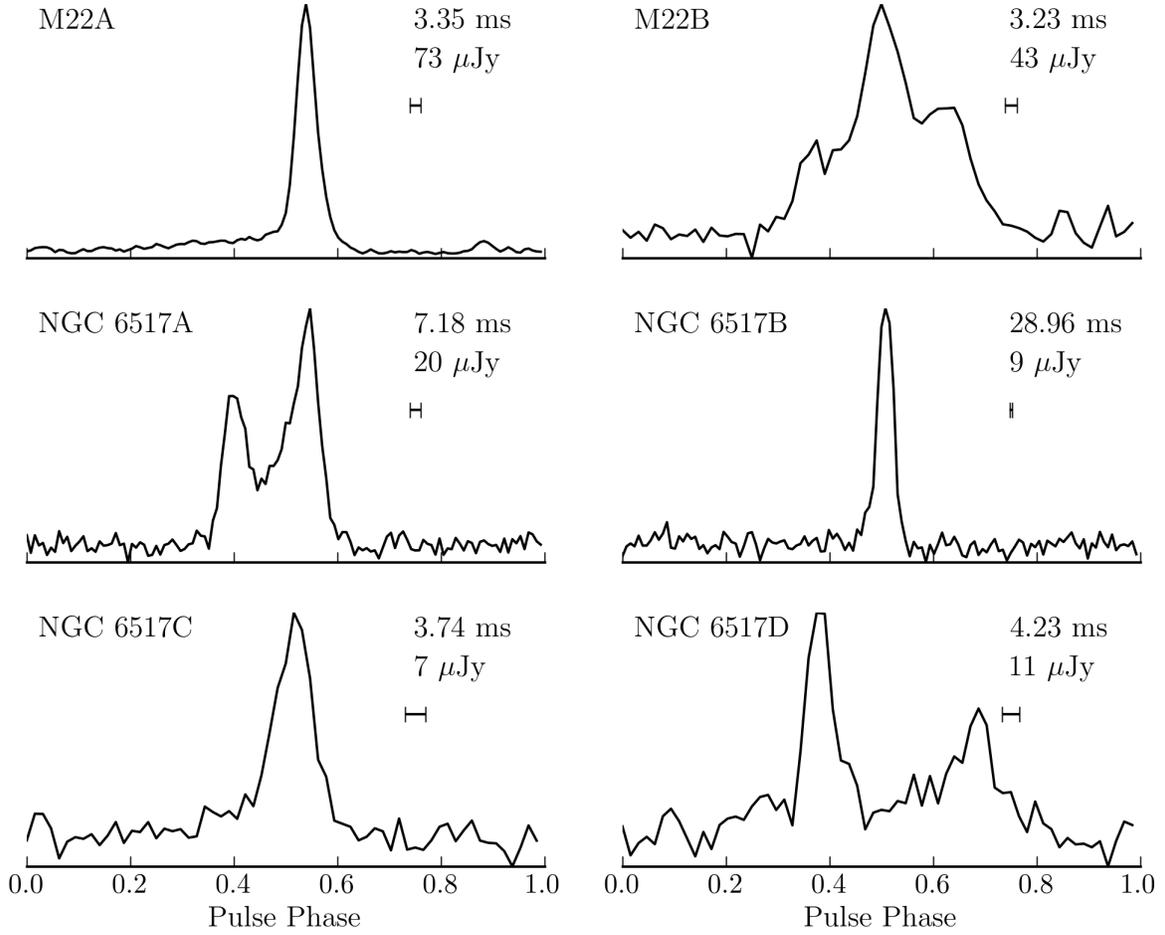}
\caption{Average pulse profiles obtained by summing all detections at
  $2\; \GHz$ for the six newly discovered pulsars.  Standard profiles
  for use in our timing analysis were created by fitting Gaussians to
  these average profiles.  Pulse periods and approximate $2\; \GHz$
  flux densities are also shown.  The horizontal bars indicate the
  contribution of dispersive smearing to the pulse width.
  \label{fig:profiles}}
\end{figure}

\begin{figure}
\centering
\includegraphics[width=6in]{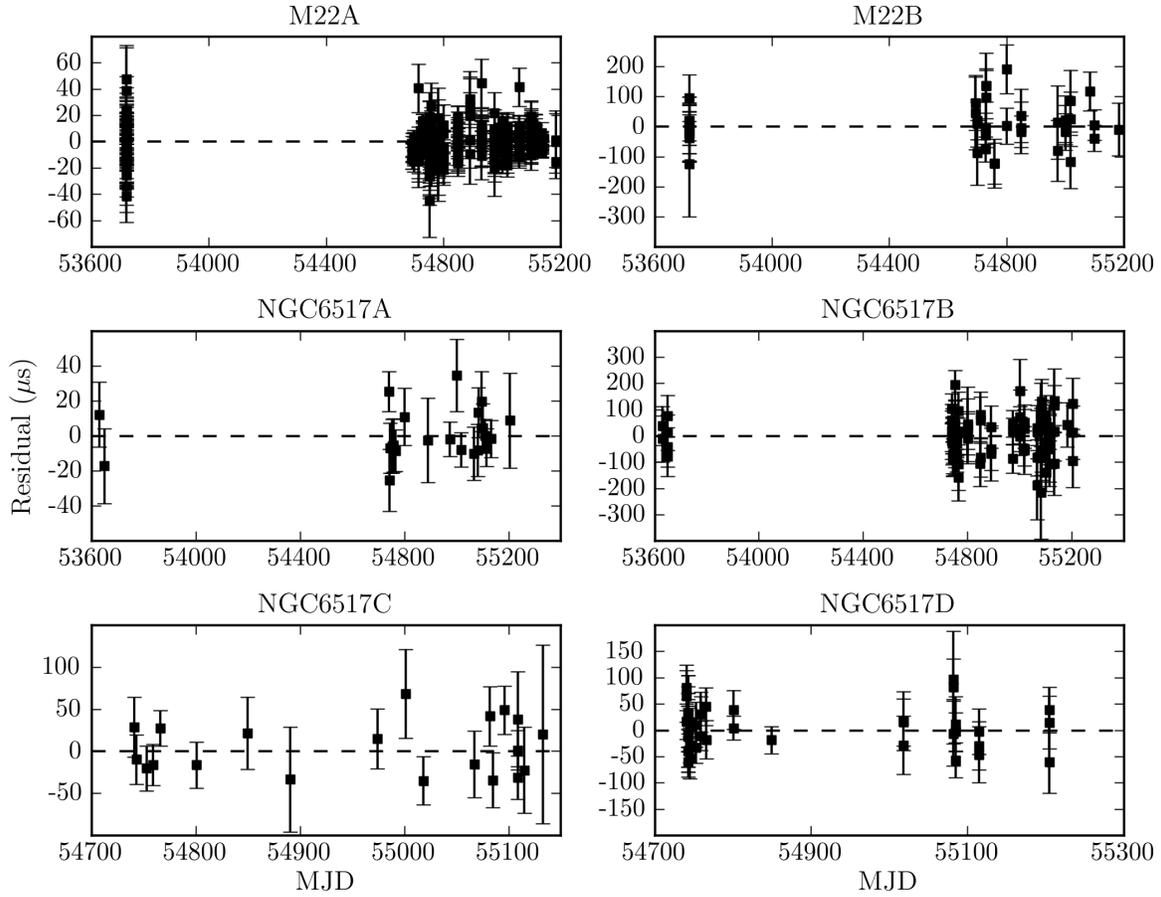}
\caption{Post-fit timing residuals for each new pulsar.  Only phase
  connected TOAs are shown.  Note the different horizontal scales in
  each panel, particularly for NGC~6517C and D, which could not be
  reliably phase connected to the discovery observations.
  \label{fig:residuals}}
\end{figure}

Six new pulsars were discovered---four in NGC~6517 and two in M22.
Timing observations for the pulsars in M22 began in 2008 August, and
follow-up of the NGC~6517 pulsars (which were discovered shortly after
those in M22) commenced in 2008 October.  Initial observations
continued to use the Spigot in the 2048 channel mode described in
\S\ref{sec:searches}, but we switched to the new Green Bank Ultimate
Pulsar Processing Instrument (GUPPI) \citep{drd+08} back-end in 2008
October.  GUPPI offers more dynamic range, improved RFI resistance,
and better sampling time ($40.96$--$64\; \us$) than the Spigot.  While
most of our timing was done at $2\; \GHz$, we obtained two $1.4\;
\GHz$ observations of M22 and one of NGC~6517.  High signal-to-noise
average pulse profiles were obtained by summing all detections for a
pulsar (see Figure \ref{fig:profiles}).  We created standard pulse
profiles by fitting one or more Gaussians to the average pulse
profiles.  Standard profiles were used to compute pulse times of
arrival (TOAs) using \texttt{PRESTO} or \texttt{PSRCHIVE}
\citep{hvm04} depending on data format.  One TOA was obtained per
observation for isolated pulsars, while multiple TOAs ($\sim 6$) were
obtained for binary pulsars to provide good sampling of the orbit.
Phase connected timing solutions were obtained using the
\texttt{TEMPO}\footnote{\url{http://tempo.sourceforge.net}} software
package and the DE405 Solar System ephemeris.  Timing solutions for
all pulsars except J1801$-$0857C and J1801$-$0857D (hereafter
NGC~6517C and D) could be reliably phase connected to the 2005
discovery observations, providing a 1574 and 1465 day baseline for the
pulsars in M22 and NGC~6517, respectively.  For NGC~6517C and D, phase
connected solutions include data spanning 463 days.  In most cases,
the reduced $\chi^2$ returned by \texttt{TEMPO} was greater than one.
Since we observe no unmodeled systematics in our fits, this is most
likely attributable to under-estimated errors on individual TOAs.  We
multiplied our TOA errors by a constant factor to obtain a reduced
$\chi^2 = 1$.  Post-fit residuals can be found in Fig.
\ref{fig:residuals}.

\section{Results \label{sec:results}}

Three of the four new pulsars in NGC~6517 are isolated MSPs
(J1801$-$0857A, hereafter NGC~6517A, NGC~6517C, and NGC~6517D).
J1801-0857B (NGC~6517B) is a partially recycled binary pulsar.  M22
contains one binary MSP (J1836$-$2354A; M22A) and one isolated MSP
(J1836$-$2354B; M22B).  The full timing solutions are given in Tables
\ref{table:isoprop} and \ref{table:binprop}, along with the offsets
from the cores of the clusters, the $2$ and $1.4\; \GHz$ mean flux
densities, and the spectral indices ($\alpha$) implied by from these
flux densities.  The coordinates of the optical centers of NGC~6517
and M22 were taken from \citet{sw86} and \citet{gra+10}, respectively.
Rough flux density measurements were made by assuming that the
off-pulse RMS noise level was described by the radiometer equation,
\begin{eqnarray}
  \sigma = \frac{T\rmsub{tot}}
                {G \sqrt{n\rmsub{pol} \Delta f\, t\rmsub{obs}}}
\end{eqnarray}
where $T\rmsub{tot}$ is the total system temperature, $G$ is the
telescope gain, $n\rmsub{pol} = 2$ is the number of summed
polarizations, and $\Delta f$ is the bandwidth.  We estimate a
10\%--20\% fractional uncertainty in these measurements, which
corresponds to an uncertainty $\alpha \pm 1.1$.  NGC~6517D was not
detected at $1.4\; \GHz$, so no information on the spectral index is
available.

Archival Chandra data exists for both clusters, but upon visual
inspection we find no sources coincident with the positions for the
pulsars obtained in our timing analysis.  NGC~6517 was observed for
$\sim 24\; \mathrm{ks}$ with ACIS-S in 2009 (Obs. ID 9597).  For a
neutral hydrogen column density $N\rmsub{H} = 3 \times 10^21\;
\pcmsq$\footnote{To calculate $N\rmsub{H}$ we used the \texttt{Colden}
  tool provided online at
  \url{http://cxc.harvard.edu/toolkit/colden.jsp}} and $D = 10.6\;
\kpc$, and assuming a limiting count rate of 10 counts, we estimate an
unabsorbed luminosity limit $L\rmsub{X} \sim 10^{32}\; \erg\, \ps$ in
the $0.3$--$8\; \keV$ band for these observations.  This limit applies
to both a $0.2\; \keV$ black body or a source with a power law
spectrum and photon index of 1.3.  M22 was observed for $\sim 16\;
\mathrm{ks}$ with ACIS-S in 2005 (Obs. ID 5437).  For $N\rmsub{H} =
1.6 \times 10^{21} \pcmsq$ and $D = 3.2\; \kpc$, we estimate
$L\rmsub{X} \sim 10^{31}\; \erg\, \ps$.  These luminosity limits lie
above the typical $L\rmsub{X} \sim 10^{30}\; \erg\, \ps$ for the MSPs
in 47 Tucanae \citep{bgh+06} and M28 \citep{b+11}, so it is likely
that any X-ray emission from the pulsars in NGC~6517 or M22 is simply
too faint to be visible given the most recent observations.

\subsection{NGC~6517B \label{sec:NGC6517B}}

The orbital period of NGC~6517B is $\sim 59~\mathrm{days}$, the fifth
longest of any known GC pulsar.  The orbit is mildly eccentric ($e =
0.03$), a trait seen in four other GC pulsars with $P\rmsub{b} >
30~\mathrm{days}$.  Any eccentricity gained during the formation of
the binary is expected to dissipate quickly, so the observed
eccentricity is likely due to gravitational perturbations of passing
stars.  Following \citet{rh95} we estimate the time required to induce
this eccentricity,
\begin{eqnarray}
  t_{>e} \simeq 4 \times 10^{11}\; \yr 
  \left (\frac{n}{10^4\; \pc^{-3}} \right )^{-1} 
  \left (\frac{v}{10\; \km\, \ps} \right ) 
  \left (\frac{P\rmsub{b}}{\mathrm{days}} \right)^{-2/3}
  e^{2/5}
\end{eqnarray}
where $n$ is the number density of stars in the neighborhood of the
pulsar, $v$ is the one-dimensional core velocity dispersion, and
$P\rmsub{b}$ is the binary period.  We were unable to find a measured
central velocity dispersion for NGC~6517 in the literature, but O.~Y.\
Gnedin
reports\footnote{\url{http://www.astro.lsa.umich.edu/~ognedin/gc/vesc.dat}}
$v = 20.6\; \km\, \ps$ based on photometric models for a mass-to-light
ratio $\Upsilon\rmsub{V} = 3\; \Msun/\mathrm{L_{\Sun,V}}$ (note that
this $\Upsilon\rmsub{V}$ is consistent with constraints we present in
\S\ref{sec:ML}).  We estimate $n$ using the luminosity density
reported in \citet{har10}, $\Upsilon\rmsub{V} = 3$ for consistency,
and an average stellar mass of $1\; \Msun$, which yields $n \approx
4.7 \times \tento{5}\; \pc^{-3}$.  This average mass is higher than
the typical $0.8\; \Msun$ turn-off mass in GCs, but is reasonable in
the cluster core, where mass segregation will have caused heavy
compact objects (stellar mass black holes, neutron stars, and massive
white dwarfs) to sink towards the center; this raises the average mass
in the neighborhood of NGC~6517B.  Using these values $t_{>e} \sim
300\; \Myr$.  This limit is directly proportional to the assumed
average stellar mass, so using a lower mass will provide a
correspondingly lower limit.

We can also place a lower limit on the characteristic age of the
pulsar by subtracting the effects of acceleration in the Galactic and
cluster gravitational potential from the observed period derivative.
The Galactic term is calculated under the approximation of a
spherically symmetric Galaxy with a flat rotation curve \citep{phi93}
and is
\begin{eqnarray}
  \dot{P}\rmsub{Gal} = -7 \times 10^{-19} \left ( \frac{P}{\s} \right )
  \left ( \cos{b} \cos{\ell} + 
    \frac{\delta - \cos{b} \cos{\ell}}
    {1 + \delta^2 - 2 \delta \cos{b} \cos{\ell}} \right )
\end{eqnarray}
where $b$ and $\ell$ are the Galactic latitude and longitude of the
cluster, $\delta = R_0/D$ and $R_0$ is the Sun's Galactocentric
distance.  \citet{phi93} showed that, to within $\sim 10\%$ accuracy
for pulsars lying less than two core radii from the cluster center,
the \emph{maximum} contribution to $\dot{P}$ from the cluster is
\begin{eqnarray}
  \dot{P}\rmsub{cluster,max} = 5 \times 10^{-12} \left (\frac{P}{\s} \right )
  \left (\frac{v}{\km\; \ps} \right )^2 
  \left [ \left (\frac{R\rmsub{c}}{\km} \right )^2 + 
    \left (\frac{R\rmsub{psr}}{\km} \right)^2 \right ]^{1/2}
\end{eqnarray} 
where $R\rmsub{c}$ is the core radius and $R\rmsub{psr}$ is the
pulsar's distance from the cluster center.  We thus find a limit on
the intrinsic spin down of $1.3 \times 10^{-17}\; \s\, \ps$, which
corresponds to $\tau\rmsub{c} \gtrsim 35\; \Myr$.  This is an order of
magnitude smaller than our estimate of $t_{>e}$, so the true value of
$\tau\rmsub{c}$ is probably higher.
Although we can only set lower limits on $t_{>e}$ and $\tau\rmsub{c}$,
they are only a few percent the age of the cluster, and much lower
than the characteristic ages of most other GC pulsars.  This raises
the possibility that NGC~6517B is young and that MSP formation has
occurred in the cluster within the last few hundred million years.

\subsection{NGC~6517D \label{sec:NGC6517D}}

NGC~6517D lies $71''$ from the cluster center, just over 20 core
radii, making it the fourth most offset GC pulsar when scaled by the
core radius ($r\rmsub{c}$).  One may question if this pulsar is truly
bound to the cluster or is just a chance alignment.  However, there
are two strong arguments for NGC~6517D actually being bound to the
cluster.  First, the pulsar is still just over one arcminute from the
cluster center, and only about one third of the tidal radius.  Second,
the DM of the pulsar is consistent with that of the other three
pulsars in NGC~6517 given the known distribution of DMs for other
bulge clusters \citep{fhn+05}, despite being $\sim 8\; \dmu$ lower
than the average of the other three MSPs.  These lines of reasoning,
when taken together, lead us to believe that NGC~6517D is likely
related to the cluster.

In that case, it is interesting to consider the probability of finding
any pulsar at such a large projected offset from the cluster center.
We begin with the assumption that the number density of pulsars as a
function of the true distance from the center is $n\rmsub{p}(r)
\propto (r^2 + r\rmsub{c}^2)^{-\alpha/2}$ \citep{phi93}.  \emph{If}
the pulsars are in thermal equilibrium with the rest of the cluster
stars, then $n\rmsub{p} \propto n\rmsub{d}^q$ where $n\rmsub{d}$ is
the number density of the dominant stars in the cluster, and $q$ is
the mass ratio of pulsars to these stars.  We do not observe the true
distance of the pulsar from the cluster center, but rather the
projected offset, $y$.  Hence, the relevant distribution is that for
the surface density, $\sigma\rmsub{p}(y) \propto (y^2
+r\rmsub{c}^2)^{-(\alpha-1)/2}$.  The number of pulsars that will lie
further out than some observed offset, $b$, is
\begin{eqnarray}
  \frac{\int_b^{r\rmsub{t}} \sigma\rmsub{p}(y)\, y\, dy}
  {\int_0^{r\rmsub{t}} \sigma\rmsub{p}(y)\, y\, dy}
\end{eqnarray}
where $r\rmsub{t}$ is the tidal radius of the cluster, which we take
to be the ``edge'' of the cluster.  NGC~6517D lies $20.7 r\rmsub{c}$
from the cluster center, and $r\rmsub{t} = 68.3 r\rmsub{c}$.  For $q =
3$ (i.e., the dominant stellar mass is $\sim 0.5\; \Msun$) we find
that only $0.01\%$ of pulsars should be found at the distance of
NGC~6517D or further.  For $q = 2$ (i.e., turn-off mass stars
dominate), this number is $3.4\%$.

In the $q = 3$ case, it seems that NGC~6517D is indeed anomalous
unless the cluster contains $\sim 10^{4}$ pulsars.  A more likely
possibility is that NGC~6517D is not in thermal equilibrium with the
rest of the stars in the cluster, as is assumed in the above analysis.
This can easily be explained if the pulsar was ejected from the
cluster core in a dynamical event.  The most likely scenario is that
the pulsar was involved in a collision with a binary (which the pulsar
may have been a member of) containing a more massive star.  Any main
sequence stars more massive than the pulsar would have died before the
progenitor of NGC~6517D, so it must have been either a massive neutron
star or a black hole.  Similar scenarios have been invoked by
\citet{ps91} to explain PSR B2127+11C in M15, and by \citet{cpg02} for
PSR J1911-5958A in NGC~6752.

\subsection{M22A \label{sec:M22A}}

M22A is the second binary pulsar in our sample and lies in a nearly
circular orbit with $P\rmsub{b} = 0.2~\mathrm{days}$.  We were unable
to measure significant orbital eccentricity, but have obtained upper
limits using two methods.  First, we have used the ELL1 binary timing
model in \texttt{TEMPO}, which was designed to fit low-eccentricity
orbits.  This model parameterizes the eccentricity and longitude of
periastron as $\epsilon_1 = e \sin{\omega}$ and $\epsilon_2 = e
\cos{\omega}$.  We used the $5\sigma$ \texttt{TEMPO} errors to compute
a conservative upper limit on the eccentricity, $e < 2.8 \times
10^{-4}$.  We also follow \citet{phi92}, computing $e\rmsub{lim} =
\delta t (a \sin{i}/c)^{-1}$ where $\delta t$ is our timing precision,
and find $e < 1.6 \times 10^{-4}$, consistent with the first analysis.

The minimum mass of the companion to M22A was calculated assuming
$M\rmsub{psr} = 1.4\; \Msun$ and $i = 90^{\circ}$ and is
$M\rmsub{c,min} = 0.017\; \Msun$, or $18$ Jupiter masses, while the
median mass corresponding to $i = 60^{\circ}$ is only $\sim 21$
Jupiter masses.  The presence of a low-mass companion in a tight orbit
makes M22A a potential new ``black widow'' pulsar, where the pulsar is
ablating its companion \citep{kbr+05}.  Many black widow pulsars
frequently eclipse due to plasma that has been stripped from the
companion.  We see no evidence for eclipses in M22A, but we believe
this is probably due to a low inclination angle so that our line of
sight never intersects any gas.  The location of M22A on the
$M\rmsub{c,min}$ vs.  $P\rmsub{orb}$ plane is similar to that of other
non-eclipsing black widow pulsars \citep[see][Fig. 1]{kbr+05}.  We are
therefore confident in classifying M22A as a black widow.

M22A is the only pulsar we have found that is bright enough for useful
polarimetry measurements.  However, we could not measure any reliable
rotation measure from our $2\; \GHz$ observations, despite searching
from $\pm 5000\; \mathrm{rad}\, \pmsq$, and see no evidence for
polarized emission.

\subsection{Cluster Mass-to-Light Ratios \label{sec:ML}}

The observed period derivative ($\dot{P}$) of GC pulsars is usually
heavily contaminated by gravitational acceleration within the cluster
potential \citep{phi93}.  This makes cluster pulsars excellent probes
of the cluster potential, and hence the enclosed mass at the projected
position of the pulsars.  Three of our newly discovered pulsars have
$\dot{P} < 0$ which, if intrinsic to the pulsars, would imply that the
pulsars are spinning \emph{up}.  This provides unambiguous evidence
that these pulsars lie on the far side of their host cluster and are
accelerating towards the Earth, and that the observed
$\dot{P}\mathrm{s}$ are dominated by the cluster potential.  This in
turn can be used to provide a lower limit on the surface mass density
within a \emph{cylinder} running through the cluster (since the true
position of the pulsar along the line-of-sight is unknown).  When
combined with the observed luminosity density, this provides a limit
on the mass-to-light ratio, $\Upsilon$.  Following \citet{dpf+02},
\begin{eqnarray}
  \Upsilon\rmsub{V} \geq 1.96 \times 10^{17}\: \dot{P}\rmsub{cluster}
  \left (\frac{P}{\s} \right )^{-1} 
  \left (\frac{\Sigma\rmsub{V}(<\theta_{\perp})}
    {10^4\; \mathrm{L_{\Sun,V}}\, \pc^{-2}} \right )^{-1}
\end{eqnarray}
where $\dot{P}\rmsub{cluster}$ signifies the contribution to
$\dot{P}\rmsub{obs}$ by gravitational acceleration in the cluster, and
$\Sigma\rmsub{V}(<\theta_{\perp})$ is the mean surface brightness
interior the position of the pulsar.  To calculate
$\Sigma\rmsub{V}(<\theta_{\perp})$, we assume a constant surface
brightness (taken from \citet{har10}) in the core of the cluster (all
the pulsars that we analyze here are within the core radius of the
cluster center).  To arrive at $\dot{P}\rmsub{cluster}$, we must
correct for other contributions to the observed $\dot{P}$, namely
\begin{eqnarray}
  \dot{P}\rmsub{cluster} = \dot{P}\rmsub{obs} - \dot{P}\rmsub{int} - 
  \dot{P}\rmsub{Gal} - \dot{P}\rmsub{pm}
\end{eqnarray}
where $\dot{P}\rmsub{int}$ is the intrinsic spin down of the pulsar
and $\dot{P}\rmsub{Gal}$ and $\dot{P}\rmsub{pm}$ are the contributions
from the potential of the Galaxy and proper motion, respectively.
Since $\dot{P}\rmsub{int}$ is unknown, we estimate it by assuming a
characteristic age of $10\; \Gyr$.  That is,
\begin{eqnarray}
  \dot{P}\rmsub{int} & \approx & \frac{P}{2\tau\rmsub{c}} \nonumber \\
  \;                 &\approx & 1.6 \times 10^{-18}\; \s^{-1}\: P .
\end{eqnarray}
The Galactic contribution was calculated as described in
\S\ref{sec:NGC6517B}.  The contribution from the proper motion of the
cluster is simply $\dot{P}\rmsub{pm} = P \mu^2 D/c$ \citep{shk70}.

The results of the above analysis for NGC~6517A, NGC~6517D, and M22B
are presented in Table \ref{table:ML}.  We were unable to find a
proper motion for NGC~6517 in the literature, but we do not expect
this to drastically change our conclusions.  For example, at the
distance of NGC~6517, a transverse velocity of $\sim 200\; \km\,
\s^{-1}$ changes our constraint on $\Upsilon\rmsub{V}$ by $< 5\%$.
The results for NGC~6517 are also particularly robust to changes in
$\dot{P}\rmsub{int}$.  In fact, if we make no corrections for
$\dot{P}\rmsub{int}$ at all, $\Upsilon\rmsub{V}$ changes by only $\sim
1\%$ for both NGC~6517A and C.  However, since M22B has a small
$\dot{P}\rmsub{obs}$, it is particularly sensitive to changes in our
assumed model for $\dot{P}\rmsub{int}$, e.g. if we assume
$\tau\rmsub{c} = 1\; \Gyr$, the limit on $\Upsilon\rmsub{V}$ increases
to $3\; \Msun/\mathrm{L_{\Sun,V}}$, but this result is still
consistent with the smaller value reported in Table \ref{table:ML}.
In all cases, we find no evidence for an anomalously high
$\Upsilon\rmsub{V}$ and our results are consistent with NGC~6517 and
M22 containing no excessive amounts of low luminosity matter.

\subsection{The Total Pulsar Content of NGC~6517
  \label{sec:NGC6517psrs}}

With the detection of four pulsars in NGC~6517 we are in a position to
say something about the total pulsar content of the cluster.  Pulsars
are observed to follow a luminosity function of the form $dN = N_0\:
L^{\alpha}\: dL$.  A commonly quoted value for $\alpha$ is $-1$,
though \citet{hrs+06} find a best-fit value of $\alpha = -0.77$ among
GC pulsars.  Given our own data, we find best-fit values of $N_0 =
2.9$ and $\alpha = -1.32$ in NGC~6517 (model A), but given the small
number of pulsars and uncertainties in the flux densities of the
pulsars and distance to the cluster, we also explore models with
$\alpha$ held fixed at $-1$ (model B) and $-0.77$ (model C).  For both
models B and C, the best-fit value for $N_0$ is about 3.0.  After
choosing appropriate bounds we may then integrate these luminosity
functions to obtain the total number of pulsars in the cluster.  We
use the luminosity of the brightest pulsar as our upper bound, and
choose a lower bound of $0.16\; \mJy\, \kpc^2$, obtained by scaling
the lowest observed GC MSP $L\rmsub{\nu}$ at $1.4\; \GHz$ ($0.3\;
\mJy\, \kpc^2$; \citep{hrs+07}) to $2\; \GHz$, assuming a spectral
index of $-1.7$.  We predict $7\: f_{\Omega}^{-1}$ to $9\:
f_{\Omega}^{-1}$ pulsars in total, where $f_{\Omega}$ is the beaming
fraction, depending on which model is used (see Table
\ref{table:Npsrs}).  It thus seems likely that we have discovered a
large fraction of the pulsars in this cluster.  However, this
conclusion does depend on the assumed value of $L\rmsub{\nu,min}$.
Since most pulsar searches are sensitivity limited, $L\rmsub{\nu,min}$
may be much lower than what is currently observed.  To illustrate the
effect of $L\rmsub{\nu,min}$ on our calculations, we also give the
total number of pulsars for $L\rmsub{2\; GHz,min} = 0.05\; \mJy\,
\kpc^2$.  This is about a factor of three lower than in the previous
calculation.  Model A, with its steeper power law index, is more
sensitive to changes in $L\rmsub{\nu,min}$.  Thus, while we cannot
rule out a significant population of low-luminosity pulsars in
NGC~6517, it seems likely that the total pulsar content is on the
order of a dozen or so.

These numbers are consistent with what we might expect based on a
simple scaling with the core interaction rate, $\Gamma\rmsub{c}$.  For
example, Terzan 5 is estimated to house $\sim 60$--$200$ pulsars
\citep{fg00} and has $\Gamma\rmsub{c} \approx 6\%$ as a fraction of
the total $\Gamma\rmsub{c}$ over all GCs.  Meanwhile, based on deep
Chandra observations, 47 Tucanae is estimated to house $\sim 25$
pulsars (independent of the radio beaming fraction) \citep{hge+05} and
has $\Gamma\rmsub{c} \approx 4\%$.  Scaling down from these numbers,
we would expect NGC~6517 to house about $12$--$17$ pulsars (if we use
the low estimate for Terzan 5).  If Terzan 5 does indeed contain over
100 pulsars, then NGC~6517 may be appear to be somewhat deficient.
However, Terzan 5 has recently been shown to contain multiple stellar
populations and is probably not a typical GC \citep{f+09}, so perhaps
it should not be surprising that it could be an outlier.  We thus find
further evidence that $\Gamma\rmsub{c}$ is a good indicator of the
pulsar content in dense clusters.

Other GCs with substantial MSP populations have been observed as
bright point sources by the Fermi Large Area Telescope (LAT)
\citep{a+10}.  We followed the technique of \citet{a+10} to see if
NGC~6517 was visible in the LAT data.  We downloaded all LAT events
within a $6^{\circ} \times 6^{\circ}$ region of interest centered on
NGC~6517 that accumulated between 7 August 2008 and 1 March 2011 (936
days).  As is common in LAT data analysis, We only selected diffuse
events and those with zenith angles $<105^{\circ}$.  We also applied a
minimum energy cut of $200\; \MeV$.  We used the standard
\texttt{P6\_V3} instrument response function and the
\texttt{gll\_iem\_v02} background model.  The Fermi point source
catalog was used to model known sources of emission, and we included
all sources within $16^{\circ}$ of NGC~6517, in order to properly
model diffuse events that came from outside our region of interest.
We then ran the \texttt{gtlike} tool, following the methods for an
unbinned likelihood analysis, and the results were used to make a test
statistic (TS) map of the region using \texttt{gttsmap}.  The TS value
around the position of NGC~6517 was only about $\sim 11$, which
corresponds to a significance $< 3.5 \sigma$.  We conclude that
NGC~6517 is not yet visible as a LAT point source, probably due to its
large distance.  However, GCs with similar numbers of MSPs have been
observed in LAT data, so it possible that NGC~6517 will be detected as
the Fermi mission continues.

\subsection{Non-detections in Five Clusters \label{sec:nondetect}}

We were unable to detect any pulsars in five of the clusters we
searched.  In light of the success of searches of clusters with
relatively large values of $\Gamma\rmsub{c}$, it is interesting to
consider what factors may have contributed to these non-detections.
In all cases, it may be that any pulsars present in these clusters lie
in tight binaries with accelerations beyond the range we searched.
Two of the most promising clusters in our sample were NGC~6388 and
Terzan~6.  NGC~6388 in particular has a higher core interaction rate
than either 47~Tucanae or Terzan~5, both of which contain dozens of
MSPs.  NGC~6388 has also been detected as a bright Fermi point source,
presumably due to the combined gamma-ray emission of an ensemble of
MSPs \citep{a+10}.  We also believed Liller~1 was a promising target
based on the Very Large Array detection of unresolved, steep-spectrum
radio emission from the cluster core, which likely arises from a
population of MSPs \citep{fg00}.  Recent analysis has also detected
Liller 1 in Fermi data \citep{tkh+11}.  We believe the most likely
explanation for our non-detections in these three clusters is
extensive scattering arising from inhomogeneities in the ISM.  All
three clusters are near the Galactic bulge and have high predicted DMs
based on the NE2001 model.  Following \citet{cor02}, we estimate the
scattering time
\begin{eqnarray}
  \log{\tau\rmsub{s}} = -3.59 + 0.129 \log{\mathrm{DM}} + 
  1.02 (\log{\mathrm{DM}})^2 - 4.4 \log{f}
\end{eqnarray}
where $\tau\rmsub{s}$ is in units of microseconds, DM is in \dmu, and
$f$ is in \GHz.  We find $\tau\rmsub{s} = 0.07, 0.25,$ and $7.2\; \ms$
for NGC~6388, Terzan~6, and Liller~1, respectively, though we note
that the typical scatter in the above relation is 0.65 in
$\log{\tau\rmsub{s}}$.  It thus seems plausible that the signal from
any MSPs in these clusters was broadened beyond the point of
detectability.  We also observed Liller~1 at $4.8\; \GHz$, hoping to
overcome scattering by taking advantage of its steep scaling with
frequency, which goes roughly as $f^{-4}$.  These searches were also
unsuccessful, probably due to a drop in pulsar flux in going to higher
frequency.

M80, on the other hand, is at a higher Galactic latitude and has a
modest predicted DM, making it similar in many ways to NGC~6517.
Also, \citet{tkh+11} report that M80 is a possible Fermi point source.
However, at $D = 10\; \kpc$, any pulsars in M80 may simply have fallen
below our limiting flux density of $11\; \uJy$.  We note that only one
MSP in NGC~6517 has flux well above this limit.  We thus conclude that
M80 lacks a bright MSP, but more sensitive searches could discover
weaker sources.  Finally, we note that NGC~6712 has the lowest
$\Gamma\rmsub{c}$ of any cluster we searched, so it is not surprising
that no new MSPs were discovered.

\section{Conclusion \label{sec:conc}}

We searched for pulsars in seven GCs with the GBT, discovering four
new pulsars in NGC~6517 and two in M22.  In both cases, these are the
first and only pulsars found in these clusters.  We were unable to
identify any of the new pulsars in archival Chandra observations.  The
binary system NGC~6517B may be only a few hundred million years old.
As this is much less then the age of the cluster, it raises the
possibility of relatively recent MSP formation in NGC~6517.  A second
binary pulsar, M22A, is a new black widow pulsar.  NGC~6517D is an
isolated MSP that lies about 20 core radii from the center of the
cluster.  It is difficult to explain the location of this pulsar
unless it underwent a dynamical encounter that ejected it from the
cluster core.  We have used the observed period derivatives of three
pulsars to constrain the mass-to-light ratio in these clusters, and in
both cases find no evidence for significant amounts of low luminosity
matter.  We have also used the observed flux densities of the pulsars
in NGC~6517 to estimate the total pulsar content of the cluster and
find that it likely houses no more than a dozen or so pulsars,
implying that we have discovered a large fraction of the population.
Despite this potentially large number of MSPs, NGC~6517 was not
detected in Fermi LAT data.  We attribute non-detections in NGC~6388,
Terzan~6, and Liller 1 to extensive scatter broadening of any pulsar
signals.  The discovery of MSPs in these systems will likely have to
wait for observing systems that offer a significant increase in
sensitivity at frequencies which are high enough to overcome
scattering effects.  M80 is similar to NGC~6517, but probably lacks an
MSP bright enough to have been detected in our searches.
 
We would like to thank an anonymous referee for helpful comments that
improved the quality of this manuscript.  R.\ S.\ Lynch was supported
by NASA grant HST-GO-10845.03A and NSF grant AST-0907967, as well as
by the GBT Student Support program.  Pulsar research at UBC is
supported by an NSERC Discovery Grant.  The National Radio Astronomy
Observatory is a facility of the National Science Foundation operated
under cooperative agreement by Associated Universities, Inc.  We
acknowldege the use of archival data from the Chandra X-ray
Observatory and Fermi LAT.

\bibliographystyle{apj}
\bibliography{ms}{}

\begin{deluxetable}{lcccccccc}
  \centering 
  \tabletypesize{\small} 
  \tablewidth{0pt} 
  \tablecolumns{9}
  \tablecaption{The Targeted High $\Gamma\rmsub{c}/D^2$ Globular Clusters \label{table:GCs}} 
  \tablehead{\colhead{ID} & 
             \colhead{$\ell$} & 
             \colhead{$b$} & 
             \colhead{$D$} & 
             \colhead{DM\tablenotemark{a}} & 
             \colhead{$\Gamma\rmsub{c}/\Gamma\rmsub{c,tot}$\tablenotemark{b}} &
             \colhead{$t\rmsub{obs}$} & 
             \colhead{$a\rmsub{max}$} &
             \colhead{$S\rmsub{nu,min}$} \\
             \colhead{} & 
             \colhead{(deg)} & 
             \colhead{(deg)} & 
             \colhead{(kpc)} & 
             \colhead{(\dmu)} & 
             \colhead{(\%)} & 
             \colhead{(hr)} & 
             \colhead{($\m\; \pssq$)} &
             \colhead{(\uJy)}}
  \startdata
  NGC~6388     & 345.56    & $-6.74$     & 9.9      & 325      & 8.0     & 1.9             & 15.4    & 38 \\
  \multirow{2}{*}{Liller 1} & \multirow{2}{*}{354.84} & \multirow{2}{*}{$-0.16$} & \multirow{2}{*}{8.2} & \multirow{2}{*}{741} & \multirow{2}{*}{1.0} & 2.8 & 3.5 & 33 \\
               &           &             &          &          &         & 5.5\tablenotemark{c} & 1.8\tablenotemark{c} & 5\tablenotemark{c} \\
  M80          & 352.67    & $+19.46$    & 10.0     & 109      & 1.7     & 2.0             & 6.9     & 11 \\
  Terzan 6     & 358.57    & $-2.16$     & 6.8      & 411      & 3.2     & 2.0             & 13.9    & 38 \\
  \bf NGC~6517 & \bf 19.23 & \bf $+6.76$ & \bf 10.6 & \bf 182.4\tablenotemark{d} & \bf 1.7 & \bf 4.7 & \bf 1.3 & \bf 7  \\
  \bf M22      & \bf 9.89  & \bf $-7.55$ & \bf 3.2  & \bf 91.4 & \bf 0.3 & \bf 1.8         & \bf 8.6 & \bf 17 \\
  NGC~6712     & 25.35     & $-4.32$     & 6.9      & 287      & 0.1     & 1.8             & 17.1    & 16 \\
  \enddata

  \tablenotetext{a}{DM estimates for all clusters except NGC~6517 and
    M22 are from the NE2001 model for the Galactic distribution of
    free electrons \citep{cl02}.}
  \tablenotetext{b}{$\Gamma\rmsub{c} \propto \rho_0^{1.5}
    r\rmsub{c}^2$, as in the text, and $\Gamma\rmsub{c,tot}$ signifies
    the sum of $\Gamma\rmsub{c}$ for all Milky Way clusters.}
  \tablenotetext{c}{These values are for the single $4.8\; \GHz$
    search of Liller 1.}
  \tablenotetext{d}{The DM of NGC~6517D was not used when calculating
    the average DM of the cluster because it is an outlier compared to
    the other three pulsars.}
  \tablecomments{Only clusters searched by R.\ S.\ Lynch are included.
    Clusters in bold face contain newly discovered pulsars. Distances,
    central densities, and core radii were all taken from
    \citet{har10}.  All maximum accelerations and limiting flux
    densities are appropriate for $3\; \ms$ pulsars with 5\% duty
    cycles and $2\; \GHz$ searches, except the second search of Liller
    1 (see note).}
\end{deluxetable}

\begin{deluxetable}{lcccc}
  \centering 
  \rotate 
  \tabletypesize{\small} 
  \tablewidth{0pt}
  \tablecolumns{5} 
  \tablecaption{Isolated Pulsars \label{table:isoprop}} 
  \tablehead{Cluster & 
             \colhead{M22} & 
             \multicolumn{3}{c}{NGC~6517}} 
  \startdata
  Pulsar Name                  & J1836-2354B           & J1801-0857A              & J1801-0857C             & J1801-0857D         \\
  Right Ascension              & 18:36:24.351(3)       & 18:01:50.6124(2)         & 18:01:50.7407(7)        & 18:01:55.3653(5)    \\
  Declination                  & $-$23:54:28.7(7)      & $-$08:57:31.85(1)        & $-$08:57:32.70(3)       & $-$08:57:24.33(3)   \\
  $\theta\rmsub{c}$\tablenotemark{a} $('')$ & 12.9     & 1.4                      & 3.4                     & 72.2                \\
  $\theta\rmsub{c}/r\rmsub{c}$ & 0.2                   & 0.4                      & 0.9                     & 20.1                \\
  $P$ (ms)                     & 3.232273969155(4)     & 7.1756146066706(8)       & 3.73869966067(2)        & 4.226532003547(3)   \\
  $\dot{P}$ (\s\, \ps)         & $-$4.8(6) \tento{-22} & $-$5.1310(4) \tento{-19} & $-$6.5(2) \tento{-20}   & 6.9(6) \tento{-21}  \\
  Reference Epoch (MJD)        & 55000                 & 54400                    & 54400                   & 54400               \\
  DM (\dmu)                    & 93.3(2)               & 182.56(1)                & 182.26(3)               & 174.71(9)           \\
  RMS Residual ($\us$)         & 64.9                  & 11.9                     & 28.1                    & 36.6                \\
  $N\rmsub{TOAs}$              & 42                    & 22                       & 22                      & 42                  \\
  $S_{2\; \GHz}\; (\uJy)$\tablenotemark{b} & 43        & 20                       & 7                       & 11                  \\
  $S_{1.4\; \GHz}\; (\uJy)$    & 40                    & 36                       & 12                      & \nodata             \\
  Spectral index, $\alpha$     & $0.20$                & $-1.6$                   & $-1.5$                  & \nodata             \\
  \enddata
  \tablenotetext{a}{Distance from the optical center of the cluster---
    $\alpha = 18$:$01$:$50.52, \delta = -08$:$57$:$31.6$ for NGC~6517
    \citep{sw86} and $\alpha = 18$:$36$:$23.94, \delta =
    -23$:$54$:$17.1$ for M22 \citep{gra+10}.}
  \tablenotetext{b}{Flux density errors are typically $\sim
    10$--$20\%$, which corresponds to an uncertainty in $\alpha$ of
    $\pm 1.1$.}
  \tablecomments{All timing solutions use the DE405 Solar System
    ephemeris and UTC(NIST) clock standard.  Reported uncertainties
    are the $1 \sigma$ \texttt{TEMPO} errors, with the errors on the
    TOAs scaled such that the reduced $\chi^2 = 1$.}
\end{deluxetable}

\begin{deluxetable}{lcc}
  \centering 
  \tabletypesize{\small} 
  \tablewidth{0pt} 
  \tablecolumns{3}
  \tablecaption{Binary Pulsars \label{table:binprop}}
  \tablehead{Cluster & 
             \colhead{M22} & 
             \colhead{NGC~6517}}
  \startdata
  Pulsar Name                       & J1836-2354A               & J1801-0857B            \\
  Right Ascension                   & 18:36:25.4452(1)          & 18:01:50.5658(5)       \\
  Declination                       & $-$23:54:52.39(3)         & $-$08:57:32.81(3)      \\
  $\theta\rmsub{c}~('')$            & 40.9                      & 1.4                    \\
  $\theta\rmsub{c}/r\rmsub{c}$      & 0.5                       & 0.4                    \\
  $P$ (ms)                          & 3.3543360829062(1)        & 28.96158773099(3)      \\
  $\dot{P}$ (\s\, \ps)              & 2.318(3) \tento{-21}      & 2.1910(5) \tento{-18}  \\
  Reference Epoch (MJD)             & 55000                     & 54400                  \\
  DM (\dmu      )                   & 89.107(2)                 & 182.39(6)              \\
  $P\rmsub{b}$ (d)                  & 0.2028278011(3)           & 59.8364526(6)          \\
  $a \sin{i}/c$ (lt-s)              & 0.0464121(6)              & 33.87545(2)            \\
  $T_0$ (MJD)                       & 54694.1962891(6)          & 54757.7226(2)          \\
  $e$                               & $< 2.8 \times 10^{-4}$    & 0.0382271(7)           \\
  $\omega$ (deg)                    & \nodata                   & 302.106(1)             \\
  Mass Function (\Msun)             & 2.6091(1) \tento{-6}      & 0.01165753(2)          \\
  $M\rmsub{c,min}$\tablenotemark{a} (\Msun) & 0.017             & 0.33                   \\
  RMS Residual ($\mu$s)             & 7.3                       & 71.2                   \\
  $N\rmsub{TOAs}$                   & 355                       & 83                     \\
  $S_{2\; \GHz}$ (\uJy)             & 73                        & 9                      \\
  $S_{1.4\; \GHz}$ (\uJy)\tablenotemark{b} & 200                & 12                     \\
  Spectral index, $\alpha$          & $-2.8$                    & $-0.81$                \\
  \enddata
  \tablenotetext{a}{We assume a pulsar mass of 1.4 \Msun.}
  \tablenotetext{b}{Flux density errors are typically $\sim
    10$--$20\%$, which corresponds to an uncertainty in $\alpha$ of
    $\pm 1.1$.}
  \tablecomments{All timing solutions use the DE405 Solar System
    ephemeris and UTC(NIST) clock standard.  Reported uncertainties
    are the $1 \sigma$ \texttt{TEMPO} errors, with the errors on the
    TOAs scaled such that the reduced $\chi^2 = 1$.}
\end{deluxetable}

\begin{deluxetable}{lccccccc}
  \centering 
  \tabletypesize{\small} 
  \tablewidth{0pt} 
  \tablecolumns{8}
  \tablecaption{Constraints on Mass-To-Light Ratio \label{table:ML}}
  \tablehead{\colhead{Pulsar}                             & 
             \colhead{$\dot{P}\rmsub{obs}$}               & 
             \colhead{$\dot{P}\rmsub{int}$}               & 
             \colhead{$\dot{P}\rmsub{Gal}$}               & 
             \colhead{$\dot{P}\rmsub{pm}$}                & 
             \colhead{$\dot{P}\rmsub{cluster}$}           &
             \colhead{$\sigma\rmsub{M}$\tablenotemark{a}} &
             \colhead{$\Upsilon\rmsub{V}$ Lower Limit}    \\
             \colhead{}                                   & 
             \multicolumn{5}{c}{($10^{-20}\; \s\, \ps$)}  &
             \colhead{($10^5\; \Msun\, \pc^{-2}$)}        &
             \colhead{($\Msun/\mathrm{L_{\Sun,V}}$)}}
  \startdata
  NGC~6517A & $-51$   & 1.2  & $-1.1$  & \nodata & $-51$   & 14   & 2.3  \\
  NGC~6517C & $-6.5$  & 0.59 & $-0.60$ & \nodata & $-6.5$  & 3.4  & 0.55 \\
  M22B      & $-0.05$ & 0.51 & $0.14$  & 0.25    & $-0.95$ & 0.57 & 0.51 \\
  \enddata
  \tablenotetext{a}{Mass surface density}
\end{deluxetable}

\begin{deluxetable}{lcccc}
  \centering 
  \tabletypesize{\small} 
  \tablewidth{0pt} \tablecolumns{5}
  \tablecaption{Total Number of Pulsars in NGC~6517
    \label{table:Npsrs}} 
  \tablehead{\colhead{Model ID} & 
             \colhead{$N_0$} & 
             \colhead{$\alpha$} & 
             \multicolumn{2}{c}{$N\rmsub{psr}\: f_{\Omega}^{-1}$ Given $L\rmsub{2\; GHz, min}$} \\
             \colhead{} & \colhead{} & \colhead{} & 
             \colhead{$0.16\; \mJy\, \kpc^2$} & 
             \colhead{$0.05\; \mJy\, \kpc^2$}} 
  \startdata
  A & 2.9 & $-1.32$ & 9  & 17 \\
  B & 3.0 & $-1.0$  & 8  & 12 \\
  C & 3.0 & $-0.77$ & 7  & 9  \\
  \enddata
  \tablecomments{For a luminosity function of the form $dN = N_0\:
    L^{\alpha}\: dL$.  Models are defined in the text.}
\end{deluxetable}

\end{document}